\documentclass[twocolumn, secnumarabic, nobibnotes, aps, prl, 
 amsmath, amssymb]{revtex4-2}

\usepackage{graphicx}
\usepackage{dcolumn}
\usepackage{bm}
\usepackage{xcolor}

\usepackage{upgreek}

\usepackage[separate-uncertainty]{siunitx}
\usepackage[breaklinks=true]{hyperref}
\hypersetup{
	colorlinks=true,
	linkcolor=blue,
	citecolor=blue,
	urlcolor=blue	
}

\begin{document}

\title{Counterpart-Induced Millisecond-Scale Truncation Mechanism of Fast Radio Bursts}

\author{Yutong He}
\email{yutong.he@mpi-hd.mpg.de}
\affiliation{Max-Planck-Institut f\"ur Kernphysik, Saupfercheckweg 1, 69117, Heidelberg, Germany}

\author{Christoph H. Keitel}
\affiliation{Max-Planck-Institut f\"ur Kernphysik, Saupfercheckweg 1, 69117, Heidelberg, Germany}

\author{Matteo Tamburini}
\email{matteo.tamburini@mpi-hd.mpg.de}
\affiliation{Max-Planck-Institut f\"ur Kernphysik, Saupfercheckweg 1, 69117, Heidelberg, Germany}

\date{\today}

\begin{abstract}
The observed millisecond-scale duration is an essential yet mysterious feature of fast radio bursts (FRBs). In this Letter, we link the observed soft gamma-ray counterpart of FRB 200428 to electron-positron pair cascades driven by Compton scattering and the Breit-Wheeler process. We demonstrate that such pair cascades can truncate FRBs to durations down to millisecond-scale, thereby establishing millisecond-scale upper bounds on their durations. The physical processes involved in the truncation mechanism occur during the propagation of FRBs after their production. Consequently, this mechanism is independent of the specific production mechanism or origin of the FRBs, suggesting that it could potentially operate in all FRBs. Our results lift the constraint on FRB production mechanisms that they must inherently generate bursts lasting only milliseconds.
\end{abstract}

\maketitle

Fast radio bursts (FRBs) represent one of the most intriguing mysteries in contemporary astrophysics~\cite{zhang.rmp.2023}. Since their discovery in 2007~\cite{lorimer.science.2007}, they have garnered growing interest with research focused on addressing fundamental questions regarding their origins and formation mechanisms~\cite{zhang.rmp.2023, platts.pr.2019, cordes.araa.2019}. Despite extensive study, many of these questions remain unanswered.

One of the key features of FRBs is their extremely short observed duration, typically of the order of milliseconds. This brief duration plays a crucial role in uncovering the origin and underlying physics of FRBs. It is widely assumed that the duration of a FRB imposes an upper limit on the size of its central engine~\cite{Zhang.nature.2020}, estimated as the product of the FRB duration and the speed of light, yielding a maximum size of $\sim10^5$~\unit{\metre}. This constraint immediately points to some of the most compact astronomical objects, such as magnetars, pulsars, or stellar-mass black holes, which have become the leading candidates for FRB sources~\cite{platts.pr.2019}. Support for a connection between FRBs and these compact objects was provided by the discovery of FRB 200428 in 2020~\cite{chime.nature.2020, bochenek.nature.2020}. This FRB enabled researchers, for the first time, to trace its origin to a specific astrophysical object, the Galactic magnetar SGR J1935+2154~\cite{mereghetti.apjl.2020, Ridnaia.na.2021, tavani.na.2021, li.na.2021}. This discovery confirmed that at least some FRBs originate from magnetars and supported the relationship between FRB duration and the maximum size of its source.

However, a recent reanalysis of observational data from FRB 20121102A revealed that this FRB produced multiple bursts on a microsecond timescale, with the shortest lasting just 6.5~\unit{\micro\second}~\cite{snelders.na.2023}. This finding challenges the assumed connection between FRB duration and source size. According to this argument, even under the most ideal conditions where the FRB source emits and then extinguishes instantaneously, a duration of 6.5~\unit{\micro\second} would still correspond to a maximum source size of $\lesssim2$~\unit{\kilo\metre}. This is significantly smaller than the possible size of magnetars, pulsars, or stellar-mass black holes~\cite{capanoNA20}. This recent discovery highlights the need to explore alternative physical processes that could govern the observed duration of FRBs. Since duration plays a critical role in proposed FRB production mechanisms, such investigations could uncover previously overlooked physics.

The proximity of the origin of FRB 200428 to Earth enabled the detection of an X-ray to soft gamma-ray counterpart both spatially and temporally coincident with the FRB, observed by multiple telescopes worldwide. This counterpart exhibits a spectrum extending to energies near $m_e c^2 \approx 0.51\text{ MeV}$, where $m_e$ is the electron mass and $c$ the speed of light in vacuum~\cite{mereghetti.apjl.2020, Ridnaia.na.2021, tavani.na.2021, li.na.2021}. Given the compact nature of the magnetar powering FRB 200428, the density of these high-energy photons in the vicinity of their source is expected to be extremely high, potentially inducing qualitatively new physical phenomena.

In this Letter, we demonstrate that an electron-positron pair cascade, driven by Compton scattering (CS) ($\gamma+e^{\pm}\rightarrow\gamma+e^{\pm}$) and the Breit-Wheeler (BW) ($\gamma+\gamma\rightarrow e^{-}+e^{+}$) process, can be initiated by the dense, energetic photons of the counterpart as they propagate through the plasma surrounding the FRB source. This cascade leads to an exponential increase in plasma density toward the rear of the counterpart photon beam, effectively impeding the propagation of both FRB and counterpart photons beyond a certain length scale. We refer to this phenomenon as the ``truncation mechanism''. For the observed parameters of FRB 200428, we find that the truncated length scale naturally corresponds to a timescale of the order of milliseconds, regardless of the initial duration of the burst. In the final paragraph of this Letter, we discuss the possibility that this mechanism may be a universal feature of FRBs, potentially explaining the short duration of both FRBs and their counterparts.

We begin by presenting a basic analytical model to illustrate the key physical mechanism, followed by more accurate kinetic simulations. For brevity, we use the term ``leptons'' to refer to electrons or positrons. We first focus on the observed counterpart photons, modeled as a collimated beam of energetic photons with around $m_e c^2$ energy, length $L$, and density $n_\gamma$, propagating through an ambient plasma with lepton density $n_e$ of initial value $n_0$, and assume $n_\gamma \gg n_e$. As the photons propagate, they scatter off leptons via CS, deviating from their initial direction of motion. These side-scattered photons no longer travel parallel to the main beam and can subsequently collide with other photons to produce leptons through the BW process. As the lepton population grows, additional photons from the original beam undergo side-scattering and convert into leptons, thereby accelerating the cascade~\cite{note}.

Without loss of generality, suppose the photon beam, initially within $-L<x<0$, propagates in the $x$ direction, with comoving coordinate $\xi \equiv x - ct$. Let the density of side-scattered photons be denoted as $n_{s}$. The continuity equations for leptons and scattered photons are given by~\cite{supp}
\begin{align}
    \partial_t n_e + \partial_{\xi} (n_e \overline{v}^e_{\xi}) & = 2 c n_\gamma \overline{\sigma}_{\text{BW}} \cdot n_{s}, \label{eq:1} \\
    \partial_t n_{s} + \partial_{\xi} (n_{s} \overline{v}^{s}_{\xi}) & = c n_\gamma  \overline{\sigma}_{\text{CS}} \cdot n_e - c n_\gamma \overline{\sigma}_{\text{BW}} \cdot n_{s}. \label{eq:2}
\end{align}
Here, $\overline{\sigma}_{\text{BW}} = \overline{\sigma}_{\text{BW}}(\xi,t)$ and $\overline{\sigma}_{\text{CS}} = \overline{\sigma}_{\text{CS}}(\xi,t)$ represent the mean cross sections of the BW and CS processes at $(\xi,t)$, respectively. Similarly, $\overline{v}^e_{\xi} = \overline{v}^e_{\xi}(\xi,t)$ and $\overline{v}^s_{\xi} = \overline{v}^s_{\xi}(\xi,t)$ denote the mean velocities of leptons and scattered photons in the $\xi$ direction at $(\xi,t)$. Note that $\overline{v}^e_{\xi}<0$ and $\overline{v}^{s}_{\xi}<0$. Since $n_\gamma \gg n_e$, this basic model neglects the depletion of the photon beam, and considers binary collisions only when they involve photons from the beam. Additionally, due to the energy threshold of the BW process, it is expected that $\overline{\sigma}_{\text{CS}} \gg \overline{\sigma}_{\text{BW}}$. In this basic model, we therefore assume that the second term on the right-hand side of Eq.~(\ref{eq:2}) is negligible (see below for kinetic simulation results). Provided that \(\overline{v}^e_{\xi}\), \(\overline{v}^{s}_{\xi}\), \(\overline{\sigma}_{\text{CS}}\), and \(\overline{\sigma}_{\text{BW}}\) can be treated as constants (see below), Eqs.~(\ref{eq:1})-(\ref{eq:2}) lead to 
\begin{equation}
    (\partial_t + \overline{v}^{e}_{\xi} \partial_{\xi})(\partial_t + \overline{v}^{s}_{\xi} \partial_{\xi}) n = 2 c^2 n_\gamma^2 \overline{\sigma}_{\text{BW}} \overline{\sigma}_{\text{CS}} \cdot n, \label{eq:3}
\end{equation}
where $n \in \{n_e, n_s\}$. Assuming $n_s = 0$ and $n_e = n_0$ at $\xi = 0$, the steady-state solution for the lepton density for $\xi < 0$ is
\begin{equation}
n_e^{\text{steady}}(\xi) = n_0  \cosh{(\sigma_{\text{eff}} n_\gamma \xi)}, \label{eq:4}
\end{equation}
with the effective cross section  
\begin{equation}
    \sigma_{\text{eff}} \equiv \left[\frac{2 \overline{\sigma}_{\text{CS}} \overline{\sigma}_{\text{BW}}}{(\overline{v}_{\xi}^{e}/c) (\overline{v}_{\xi}^{s}/c) } \right]^{1/2}. \label{eq:5}
\end{equation}
This solution indicates that the lepton density increases exponentially toward the rear of the photon beam. To examine the temporal evolution of $n_e$, we note that frequent collisions result into $\overline{v}^{e}_{\xi} \sim \overline{v}^{s}_{\xi} \sim c$ (see~\cite{Faure.pre.2024} and the kinetic simulation results discussed below). Introducing the mean velocity $v_* \equiv -(\overline{v}^{e}_{\xi}\overline{v}^{s}_{\xi})^{1/2}$, we approximate Eq.~(\ref{eq:3}) by replacing $\overline{v}^{e}_{\xi}$ and $\overline{v}^{s}_{\xi}$ with $v_*$. Since the operator $(\partial_t + v_* \partial_{\xi})$ is a material derivative, this approximation implies that Eq.~(\ref{eq:3}) describes the exponential evolution of a density element $n$ drifting with velocity $v_*$. Accordingly, we expect that the ambient plasma at the leading edge of the photon beam ($\xi = 0$) induces an exponential increase in $n_e$, with a peak receding at an approximate velocity $v_*$.

The pair cascade converts beam photons into leptons and scattered photons, resulting in the depletion of the energetic photon beam.  According to Eq.~(\ref{eq:4}), this depletion occurs over a characteristic length scale $L_{\text{cut}}^{\text{CP}} \sim \ln \left(2 n_\gamma / n_0\right) / \left(\sigma_{\text{eff}} n_\gamma\right)$. Notably, due to its logarithmic dependence, $L_{\text{cut}}^{\text{CP}}$ is only weakly sensitive to even orders-of-magnitude variations in $n_\gamma/n_0$, implying that $\sigma_{\text{eff}}$ and $n_\gamma$ are the primary parameters governing the truncation length scale. Assume $n_0 \ll n_c \ll n_\gamma$ with $n_c \sim 10^{15\text{--}17}~\unit{\per\cubic\metre}$ being the critical density for FRB photons (corresponding to frequencies from a few hundred MHz to several GHz~\cite{chime.nature.2019, Gajjar.aj.2018}). In the absence of a strong background magnetic field, the plasma becomes opaque to FRB photons at a distance $L_{\text{cut}}^{\text{FRB}}$ where $n_e \gtrsim n_c$. Owing to the exponential growth of $n_e$, the lepton density rapidly approaches $n_\gamma$ so that $L_{\text{cut}}^{\text{FRB}}\sim L_{\text{cut}}^{\text{CP}}$. By contrast, in environments with strong background magnetic fields (e.g., the magnetosphere of a magnetar), FRB photons may propagate as extraordinary modes through plasmas with densities exceeding $n_c$~\cite{zhang.rmp.2023, swanson.plasma_waves}. However, since the magnetic field strength decreases rapidly with distance, the cascade-driven increase in $n_e$ prevents FRB photon propagation once the local lepton gyrofrequency falls below the photon frequency (see Supplemental Material (SM)~\cite{supp}).

Considering that the momentum dependence of the CS and BW cross sections reduces their values below the Thomson cross section $\sigma_{T} \approx 6.65 \times 10^{-29}~\unit{\square\metre}$, we estimate $\sigma_{\text{eff}} \sim 10^{-30}~\unit{\per\cubic\metre}$ (see below). The observed counterpart of FRB 200428 enables an estimate of the photon density near the source as $n_\gamma \sim n_\gamma^{\text{Earth}} D_{\text{source}}^2 / R^2 \sim 10^{23\text{--}29}~\unit{\per\cubic\metre}$, where $n_\gamma^{\text{Earth}} \sim 10^{-4}~\unit{\per\cubic\metre}$ is the photon density of the counterpart observed from Earth near the $m_e c^2$ energy range~\cite{Ridnaia.na.2021, li.na.2021, supp}. Here, $D_{\text{source}} \sim10~\text{kiloparsecs (kpc)}\approx 3\times10^{20}~\unit{\metre}$ denotes the distance from the magnetar SGR J1935+2154 to Earth~\cite{bailes.mnras.2021, Zhou.apj.2020, Zhong.apjl.2020, kothes.apj.2018}. Based on the recent identification of the emission radius of FRB 20221022A~\cite{nimmo.nature.2025}, we assume $R \sim 10^{4\text{--}7}~\unit{\metre}$ is the radius of the region from which the counterpart is emitted. The ambient plasma density is estimated to lie in the range $n_0 \sim 10^{6\text{--}19}~\unit{\per\cubic\metre}$, with the upper bound corresponding to the Goldreich-Julian~\cite{goldreichAJ69} density at the surface of SGR J1935+2154~\cite{israel.2016.mnras}. This yields a truncation timescale $L_{\text{cut}}^{\text{CP}}/c$ of $10^{-3}\text{--}10^{3}~\unit{\milli\second}$, which is consistent with the observed durations of FRBs. For instance, the two radio components of FRB 200428 have durations of 0.6~\unit{\milli\second} and 0.3~\unit{\milli\second}~\cite{chime.nature.2020}. Furthermore, the collisional frequency of leptons and scattered photons with the beam photons can be estimated as $\nu \sim n_\gamma \sigma_T c$. This implies that the four kinematic parameters $\overline{v}^e_{\xi}$, $\overline{v}^{s}_{\xi}$, $\overline{\sigma}_{\text{CS}}$, and $\overline{\sigma}_{\text{BW}}$ reach their steady-state values after a characteristic length (time) of the order of $L_{\text{steady}} \sim c/\nu$ ($\nu^{-1}$). Since $L_{\text{steady}} \ll L_{\text{cut}}^{\text{FRB}} \sim L_{\text{cut}}^{\text{CP}}$, our previous approximation of treating $\overline{v}^e_{\xi}$, $\overline{v}^{s}_{\xi}$, $\overline{\sigma}_{\text{CS}}$, and $\overline{\sigma}_{\text{BW}}$ as constants is well justified.

Motivated by these estimates, we conducted a more accurate quantitative investigation. To simulate the system, we implemented the CS and BW processes into the particle-in-cell (PIC) code EPOCH~\cite{arber.2015.epoch}. Details and benchmarking results are provided in the SM~\cite{supp}. The exponential growth in the number of pairs results in an orders-of-magnitude increase in the number of simulation particles, along with significantly higher spatial and temporal resolution requirements. Fortunately, the collimation of the photon beam and the rapid development of the cascade allow the cascade dynamics to be captured by considering only the beam’s propagation direction in space. Accordingly, we performed 1D3V PIC simulations solely to confirm the system's dynamics as predicted by the basic model and to determine the characteristic value of $\overline{v}^e_{\xi}$, $\overline{v}^{s}_{\xi}$, $\overline{\sigma}_{\text{CS}}$, and $\overline{\sigma}_{\text{BW}}$. These parameters were subsequently used to predict the truncation time by numerically solving a set of governing equations that incorporate 3D geometric effects and photon beam depletion (see below). Notably, the numerical solution of the 3D governing equations, using the 1D3V-PIC-derived parameters as constants, validated our basic model by yielding truncation times of the same order of magnitude as predicted by the analytical model.

\begin{figure}
\includegraphics[width=\linewidth]{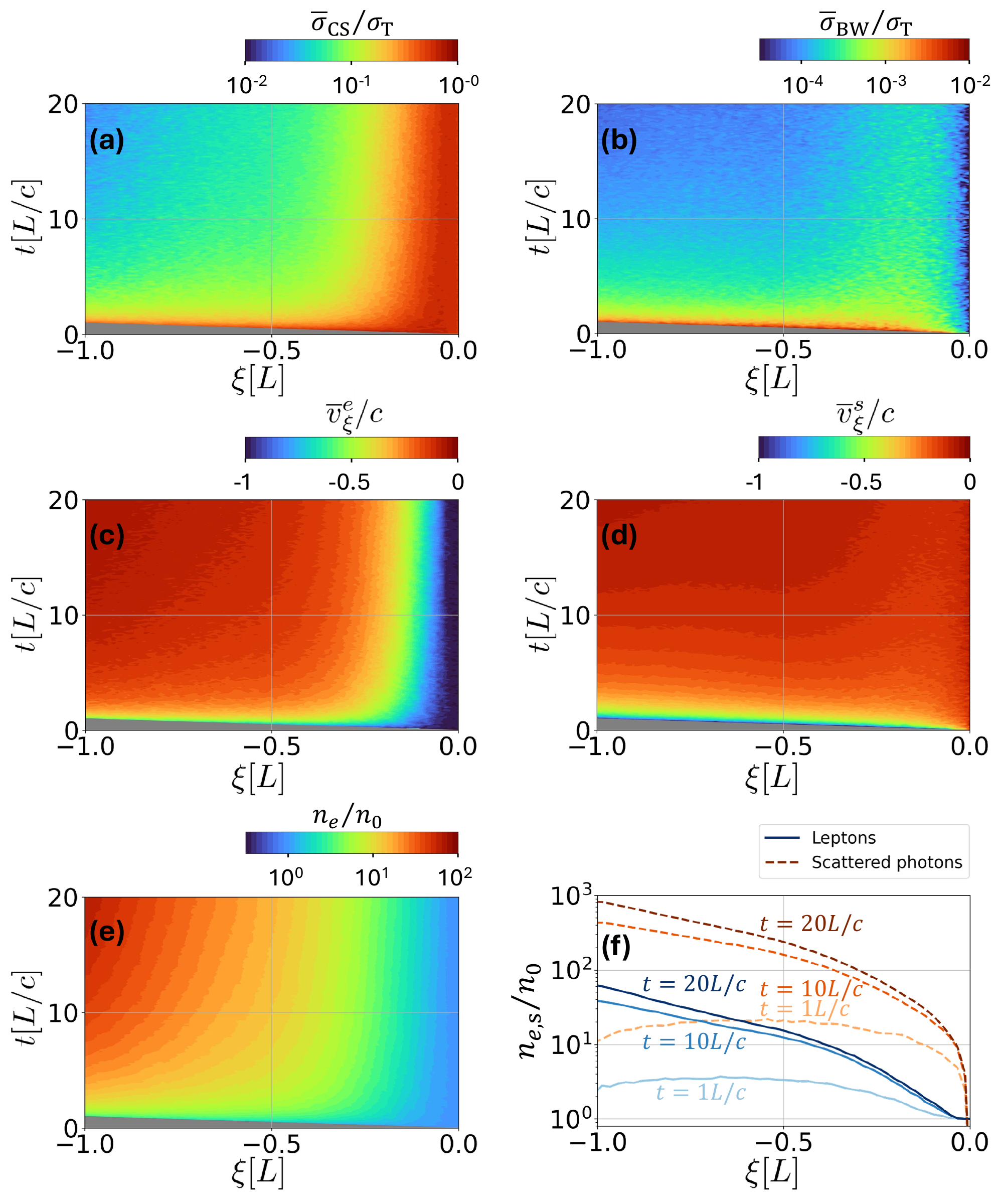}
\caption{1D3V PIC simulation results for $L = 10^{5.5}$~\unit{\metre}  and $n_\gamma = 10^{24.5}$~\unit{\per\cubic\metre}.
(a)-(b) Contour plot of the spatial and temporal evolution of the mean cross sections $\overline{\sigma}_{\text{CS}}$ and $\overline{\sigma}_{\text{BW}}$, normalized by the Thomson cross section $\sigma_T$.
(c)-(d) Contour plot of the spatial and temporal evolution of the mean drift velocities $\bar{v}^e_\xi$ and $\bar{v}^s_\xi$, normalized by $c$.
(e) Contour plot of the spatial and temporal evolution of the lepton density $n_e$, normalized by the initial plasma density $n_0$.
(f) Snapshots of the lepton density $n_e$ (solid lines) and scattered photon density $n_s$ (dashed lines), normalized by $n_0$, at $t = 1L/c$, $10L/c$, and $20L/c$.} \label{fig:1}
\end{figure}

Equation~(\ref{eq:4}) indicates that the development of the cascade is governed by the critical parameter $\kappa \equiv \sigma_{\text{eff}} n_\gamma L$, with the cascade occurring when $\kappa$ exceeds unity. Consistent with the above estimated counterpart photon density of FRB 200428 near its source, the PIC simulation is initialized with a photon beam of density $n_\gamma = 10^{24.5}~\unit{\per\cubic\metre}$ and length $L = 10^{5.5}~\unit{\metre}$, yielding $\kappa \approx 1$ to ensure the development of a moderate cascade. The beam is initialized within the spatial region $-L < x < 0$ and propagates in the positive $x$ direction, where it interacts with a tenuous ambient electron-proton ($e^{-}-p$) plasma of density $n_0 = 10^{9}~\unit{\per\cubic\metre}$ located in the region $x > 0$. Higher plasma densities, as well as a dominance of leptons over ions, are expected if the counterpart is generated within or near a magnetar’s magnetosphere~\cite{goldreichAJ69} (see below for PIC results involving an ambient pair plasma). In general, according to Eq.~(\ref{eq:4}), an increase in $n_0$ is expected to mildly shorten the cascade development time. We run the simulations over the interval $0 < t < 20L/c$, where simulation results confirm that $\overline{v}^e_{\xi}$, $\overline{v}^{s}_{\xi}$, $\overline{\sigma}_{\text{CS}}$, and $\overline{\sigma}_{\text{BW}}$ reach their steady value [Fig.~\ref{fig:1}(a)-(d)], and the profile of $n_{e}$ stabilizes [Fig.~\ref{fig:1}(f)]. Based on the observed counterpart spectrum reported in Refs.~\cite{mereghetti.apjl.2020, Ridnaia.na.2021, tavani.na.2021, li.na.2021}, the initial photon population is sampled according to a power-law distribution $dn_\gamma/d\varepsilon \propto \varepsilon^{-3}$ from 200~keV to 3~MeV~\cite{supp}.

The resulting pair cascade is illustrated in Fig.~\ref{fig:1}. Panels~(a)-(d) depict the spatial and temporal evolution of the four parameters: $\overline{v}^e_{\xi}$, $\overline{v}^{s}_{\xi}$, $\overline{\sigma}_{\text{CS}}$, and $\overline{\sigma}_{\text{BW}}$. Panel~(e) shows the spatial and temporal evolution of the lepton density $n_e$, confirming its exponential growth toward the rear of the photon beam. As illustrated in Fig.~\ref{fig:1}(f), the exponential growth rates of $n_e$ and $n_s$ with respect to $\xi$ stabilize at approximately $\xi \approx -0.4L$, which is consistent with the stabilization of the four parameters around the same location, as shown in panels~(a)-(d). Additionally, these results confirm that $\overline{\sigma}_{\text{CS}} n_{e} \gg \overline{\sigma}_{\text{BW}} n_{s}$ by approximately one order of magnitude, thereby justifying our previous assumption that the BW contribution could be neglected in Eq.~(\ref{eq:2}).  

\begin{figure}
\includegraphics[width=\linewidth]{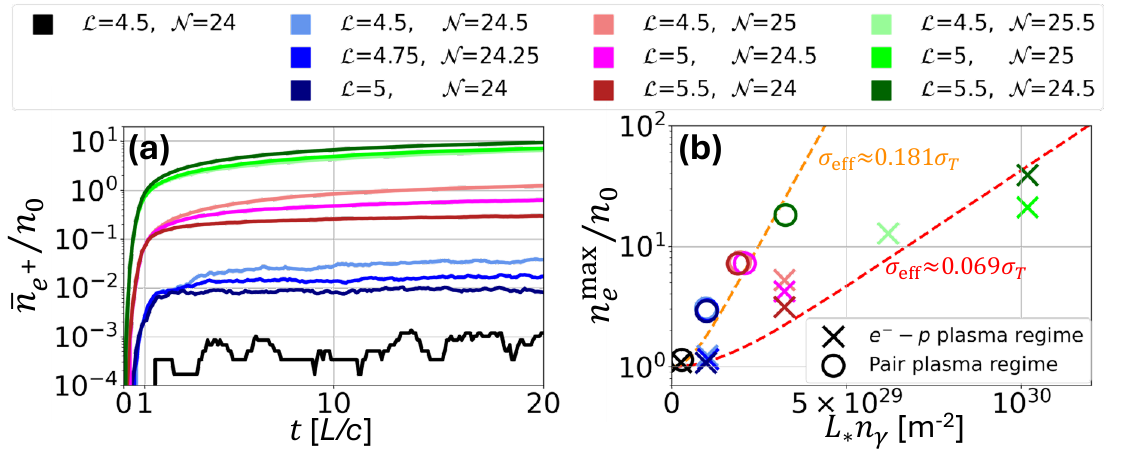}
\caption{1D3V PIC simulation results. (a)~Temporal evolution of the mean positron density $\overline{n}_{e^{+}}$ averaged over the range $-L<\xi<0$ and normalized by the initial plasma density $n_0$. (b)~Scaling of the normalized peak lepton density $n^{\text{max}}_e/n_0$ as a function of $L_* n_{\gamma}$. In the legend $\mathcal{L} \equiv \log_{10}(L[\unit{\metre}])$ and $\mathcal{N} \equiv \log_{10}(n_\gamma[\unit{\per\cubic\metre}])$.} \label{fig:2}
\end{figure}

To further test the prediction that the development of the cascade is primarily determined by $\kappa$, a set of PIC simulations was conducted, varying $n_\gamma$ and $L$. Specifically, simulations were performed for $n_\gamma L [\unit{\per\square\metre}] = 10^{28.5}$, $10^{29}$, $10^{29.5}$, and $10^{30}$ to examine the transition from $\kappa \lesssim 1$ to $\kappa \gtrsim 1$. For each of the latter three values of $n_\gamma L$, we considered three different combinations of $L$ and $n_\gamma$, which are listed in the legend at the top of Fig.~\ref{fig:2}. 

Panel~(a) of Fig.~\ref{fig:2} plots the temporal evolution of the positron density $\overline{n}_{e^+}$ averaged over the range $-L < \xi < 0$. The results demonstrate convergence to a steady value within a few multiples of $L/c$, even in systems where only a small number of pairs are produced. Moreover, Fig.~\ref{fig:2}(a) shows that pair cascades develop ($\overline{n}_{e^+}/n_0 \gg 1$) only if the system satisfies $\kappa \gtrsim 1$, as displayed by the three green lines. Conversely, when $\kappa < 1$, the system produces an insignificant number of pairs ($\overline{n}_{e^+}/n_0 \ll 1$). 

To estimate $\sigma_{\text{eff}}$, we determine the position $L_*$ such that $n_e(\xi = -L_*) = n_e^{\text{max}}$ at $t = 10L/c$, where $n_e^{\text{max}}$ is the maximum lepton density within the interval $-L < \xi < 0$. The interval $-L_* < \xi < 0$ thus approximately corresponds to the region where the cascade has developed. Figure~\ref{fig:2}(b) illustrates the scaling of $n_e^{\text{max}}$ with respect to $L_* n_\gamma$. According to Eq.~(\ref{eq:4}), the red dashed line in Fig.~\ref{fig:2}(b) represents a fitted function of the form $\cosh{(\sigma_{\text{eff}} n_\gamma L_*)}$, where $\sigma_{\text{eff}}$ is the fitting parameter. The best-fit value for the effective cross-section is found to be $\sigma_{\text{eff}} \approx 0.069\sigma_T$. Notably, panels ~(a) and (b) of Fig.~\ref{fig:2} show that systems with the same value of $n_\gamma L$ exhibit similar values of $\overline{n}_{e^+}/n_0$ and $n_e^{\text{max}}/n_0$. This result corroborates the prediction of our model that pair creation in the system is primarily governed by $\kappa$. 

To investigate the pair plasma regime, we conducted the same set of PIC simulations as described above but initialized the ambient plasma with electrons and positrons. The circular markers in Fig.~\ref{fig:2}(b) represent the results for $n_{e}^{\text{max}}/n_0$ in this regime, while the orange dashed line corresponds to the prediction based on the same fitting of the red dashed line, using the best-fit value for the effective cross-section: $\sigma_{\text{eff}} \approx 0.181\sigma_T$. The increase in this fitted value of $\sigma_{\text{eff}}$ compared to that obtained for an electron-proton plasma ($\sigma_{\text{eff}} \approx 0.069\sigma_T$) arises because, in the absence of the restoring force exerted by protons, the forward momentum imparted by photons induces a collective bulk motion of leptons. This motion causes $\overline{v}^e_{\xi}$ and $\overline{v}^{s}_{\xi}$ to approach zero, leading to a higher effective cross-section $\sigma_{\text{eff}}$ [see Eq.~(\ref{eq:5})]. The two fitted values of the effective cross-section, approximately $4.5 \times 10^{-30}$~\unit{\square\metre} and $1.2 \times 10^{-29}$~\unit{\square\metre}, are consistent with the value ($10^{-30}$~\unit{\square\metre}) adopted for the basic model estimate of the truncation time.

Thus far, we have neglected the decrease in particle density with increasing distance $r$ from the source in three-dimensional space. Additionally, the CS and BW processes deplete the photon beam. To account for these effects, we consider the following system of continuity equations in spherical coordinates~\cite{supp}
\begin{align}
    & \partial_t (r^2 n_e) + \overline{v}^e_{\zeta} \partial_\zeta (r^2 n_e) = 2 c \overline{\sigma}_{\text{BW}} r^2 n_{s} n_\gamma \label{eq:7} \\
    & \partial_t (r^2 n_{s}) + \overline{v}^s_{\zeta} \partial_\zeta (r^2 n_{s}) = c [\overline{\sigma}_{\text{CS}}n_{e} - \overline{\sigma}_{\text{BW}} n_{s}]r^2 n_\gamma \label{eq:8} \\
    & \partial_t (r^2 n_\gamma) = -c [\overline{\sigma}_{\text{BW}} n_{s} + \overline{\sigma}_{\text{CS}} n_e] r^2 n_\gamma. \label{eq:9}
\end{align}
Here $\zeta \equiv r - ct - R_0$, where we assume that the energetic photons originate from $r = R_0$, with $r = 0$ corresponding to the center of the source. Note that $r = r(\zeta, t) = \zeta + ct + R_0$ is a function of both $\zeta$ and $t$. The mean velocities of leptons and scattered photons in the $\zeta$ coordinate are denoted by $\overline{v}^e_{\zeta}$ and $\overline{v}^s_{\zeta}$, respectively. Figure~\ref{fig:2}(a) shows that the cascades develop within a few times $L/c$, indicating that FRB and counterpart truncation occur on a similar timescale. We thus estimate $\overline{v}^e_{\zeta}$, $\overline{v}^s_{\zeta}$, $\overline{\sigma}_{\text{CS}}$, and $\overline{\sigma}_{\text{BW}}$ by averaging the simulation results from Fig.~\ref{fig:1}(a-d) over $0 < t < 2L/c$ and $-L < \xi < 0$, obtaining $\overline{v}^e_{\zeta} \approx -0.55c$, $\overline{v}^s_{\zeta} \approx -0.62c$, $\overline{\sigma}_{\text{CS}} \approx 0.33\sigma_T$, and $\overline{\sigma}_{\text{BW}} \approx 0.0023\sigma_T$. These values support the model assumption $v_* \sim c$, and yield an effective cross section $\sigma_{\text{eff}} \approx 0.067\sigma_T$, consistent with effective cross sections reported in Fig.~\ref{fig:2}(b). The boundary conditions specify that, for all $t$, $n_\gamma = n_\gamma^0$ at $r = R_0$, where $n_\gamma^0$ is the photon beam density at the emission surface $R_0$. Additionally, the leading edge of the photon beam, located at $\zeta = 0$ (initially at $R_0$), enters a uniform plasma with $n_e = n_0$ and $n_s = 0$. The precise values of $R_0$ and $n_\gamma^0$ remain uncertain. To address this, we numerically solve Eqs.~(\ref{eq:7})--(\ref{eq:9}) for $n_e$, $n_s$, and $n_\gamma$, and perform a parameter scan over $R_0$ and $n_\gamma^0$.

\begin{figure}
\includegraphics[width=\linewidth]{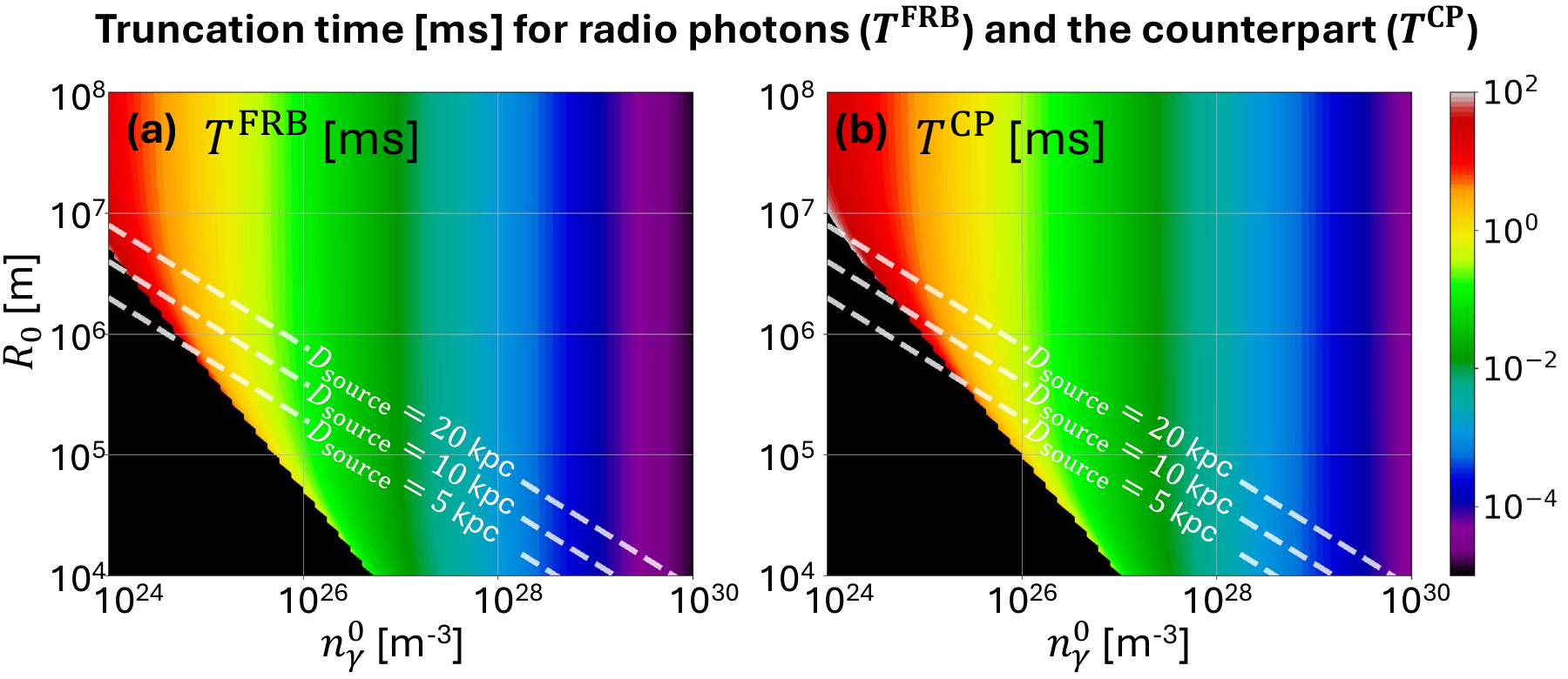}
\caption{Truncation time (color scale, in milliseconds) for (a) the FRB ($T^{\text{FRB}}$) and (b) the counterpart ($T^{\text{CP}}$), obtained by numerically solving Eqs.~(\ref{eq:7})--(\ref{eq:9}) as a function of $R_0$ (the emission site radius) and $n_\gamma^0$ (the local photon density at $r = R_0$). White dashed lines correspond to the scaling relation $n_\gamma^0 R_0^2 \approx n_\gamma^{\text{Earth}} D_{\text{source}}^2$ for three estimated distances $D_{\text{source}}$ from the origin of FRB 200428 to Earth. The black regions in the lower-left corners of both panels denote parameter regions in which truncation does not occur.} \label{fig:3}
\end{figure}

The results of this analysis are presented in panels~(a) and (b) of Fig.~\ref{fig:3}, which show the truncation times for FRB photons $T^{\text{FRB}}$ and for the counterpart $T^{\text{CP}}$, respectively. Here, $T^{\text{FRB}}$ is defined as the duration required for $n_e$ to reach the critical density for FRB photons, taken to be $n_c = 10^{16}~\unit{\per\cubic\metre}$ (corresponding to a photon frequency of approximately 900~MHz). Similarly, $T^{\text{CP}}$ is defined as the duration at which $n_\gamma$ decreases to 1\% of the value it would have in the absence of the CS and BW processes. When $n_\gamma \sim n_s \sim n_e$, additional processes--such as CS between scattered photons and leptons, BW pair production between scattered photons, and electron-positron annihilation into photons--may affect $T^{\text{CP}}$ (FRB truncation is expected to have already occurred at $n_e \sim n_c \ll n_\gamma$). To evaluate their impact, we simulated these processes with the PIC code~\cite{supp}, and verified that they have a minor effect on photon beam depletion under the conditions considered here. This is attributed to the fact that these processes do not alter the total number of particles that can collide with the photon beam. Motivated by the recent discovery of the emission radius of FRB 20221022A~\cite{nimmo.nature.2025}, we perform a parameter scan over the range $10^4~\unit{m} < R_0 < 10^8~\unit{m}$, and estimate $n_{\gamma}^0$ using the scaling relation $n_\gamma^0 R_0^2 \approx n_\gamma^{\text{Earth}} D_{\text{source}}^2$. Based on the observed counterpart spectrum, we adopt $n_\gamma^{\text{Earth}} = 1.66 \times 10^{-4}~\unit{\per\cubic\metre}$~\cite{supp}. Reported values for $D_{\text{source}}$ range from a few kpc~\cite{bailes.mnras.2021, Zhou.apj.2020, Zhong.apjl.2020} to over ten kpc~\cite{kothes.apj.2018}. The corresponding region in the $(n_\gamma^0, R_0)$ parameter space is highlighted in Fig.~\ref{fig:3} using white dashed lines, with $D_{\text{source}}$ varying from 5 to 20~kpc.

Figure~\ref{fig:3} confirms that within the region bounded by the white dashed lines, the truncation times are up to the milliseconds scale. Moreover, the truncation time is primarily governed by the local photon density $n_\gamma^0$, in agreement with our basic model and PIC simulation results. Specifically, the model predicts that a cascade develops when $\kappa\equiv\sigma_{\text{eff}}n_{\gamma}L_\kappa \gtrsim 1$, and our simulations indicate that $\sigma_{\text{eff}}$ depends only weakly on the specific parameter choices. As a result, the characteristic length scale $L_\kappa$ over which $\kappa \sim 1$ is mainly determined by $n_\gamma^0$. This supports the robustness of the model and suggests that, once initiated, the cascade develops over a relatively short distance, such that the reduction in particle density due to spatial expansion away from the source has a relatively minor effect. However, the results in Fig.~\ref{fig:3} also indicate that if $n_\gamma^0$ is too low, or if the photons are emitted too close to the source--resulting in a rapid decrease in photon density--truncation may not occur.

To summarize, we have identified a cascade process capable of truncating the duration of FRBs and their counterparts to millisecond or even shorter timescales. This implies that the emission mechanism responsible for FRBs need not be intrinsically constrained to produce short bursts. The proposed mechanism is informed by observational data from FRB 200428 and its associated high-energy counterpart. While such counterparts have not been observed for all FRBs, we emphasize that this truncation mechanism may operate universally. The detection of a counterpart in FRB 200428 is primarily attributed to its proximity, as it is both the closest FRB observed to date and the only one detected within our galaxy. Assuming other FRBs possess counterparts with similar radio-to-X-ray luminosity ratios, their counterparts would likely remain undetectable due to their much greater distances~\cite{zhang.rmp.2023}. This suggests that similar high-energy counterparts may be common among FRBs, even if they have not yet been observed.

Although the physical conditions in the emission regions of FRBs and their counterparts are not fully understood and may be more complex than those considered here, we expect the fundamental physical processes governing cascade development to remain unchanged. This expectation is supported by a parameter scan conducted over an intentionally broad and exaggerated range of  $\overline{v}^e_{\zeta}$, $\overline{v}^s_{\zeta}$, $\overline{\sigma}_{\text{CS}}$, and $\overline{\sigma}_{\text{BW}}$, which demonstrates that the predicted truncation times are relatively insensitive to variations in these parameters (see SM~\cite{supp}).

The nature of this mechanism, where truncation time is primarily governed by the local energetic photon density, also makes it consistent with observations of ultrashort, microsecond-scale FRBs~\cite{snelders.na.2023}. Furthermore, it may account for the rarity of FRB-soft gamma repeater associations~\cite{linN20}. In cases where the local photon density is excessively high, FRB signals may be truncated to durations too short to be detected, or the counterpart-induced cascade may develop ahead of the FRB photons, entirely preventing their propagation. 

\bibliography{apssamp}

\begin{thebibliography}{28}
\bibitem [{not()}]{note}%
  \BibitemOpen
  \href@noop {} {}\bibinfo {note} {Photon-proton scattering is subdominant, its
  cross section being suppressed due to the proton's large mass. Moreover, as
  the cascade develops, leptons rapidly outnumber protons. Bethe-Heitler and
  other higher-order processes are suppressed by the fine structure constant
  $\alpha \approx 1/137$.}\BibitemShut {Stop}%
\end{thebibliography}

\end{document}